\documentclass[showpacs,preprintnumbers,aps]{revtex4}
\usepackage{amsfonts}
\usepackage{amssymb}
\usepackage{amsmath}
\usepackage[all]{xy}
\begin{document}
\preprint{USM-TH-189}
\newcommand{\ba}{\begin{eqnarray}}
\newcommand{\ea}{\end{eqnarray}}
\newcommand{\be}{\begin{equation}}
\newcommand{\ee}{\end{equation}}
\newcommand{\bib}{\bibitem}
\newcommand{\ed}{\end{document}}
\newcommand{\nn}{\nonumber\\}
\newcommand{\fr}{\frac}
\newcommand{\wt}{\widetilde}
\title{Confinement from constant field condensates}
\author{Patricio Gaete}
\email {patricio.gaete@usm.cl} \affiliation{Departamento de
F\'{\i}sica, Universidad T\'ecnica F. Santa Mar\'{\i}a,
Valpara\'{\i}so, Chile}
\author{Eduardo Guendelman}
\email {guendel@bgumail.bgu.ac.il}
\affiliation{Physics Department, Ben Gurion University, Beer
Sheva 84105, Israel}
\author{Euro Spallucci}
\email {euro@ts.infn.it} \affiliation{Dipartimento di Fisica
Teorica, Universit\`a di
Trieste and INFN, Sezione di Trieste, Italy}
\date{\today}

\begin{abstract}
For a $(2+1)$-dimensional reformulated $SU(2)$ Yang-Mills theory,
we compute the interaction potential within the framework of the
gauge-invariant but path-dependent variables formalism. This
reformulation is due to the presence of a constant gauge field
condensate. Our results show that the interaction energy contains
a linear term leading to the confinement of static probe charges.
This result is equivalent to that of the massive Schwinger model.
\end{abstract}
\maketitle

\section{Introduction}

One of the most important problems facing $QCD$ in the infrared
region is the confinement of quarks. One expects that the gluon
field exchanged between a quark and an antiquark gathers up into a
string, thus preventing the quarks from separating to large
distances. As is well known, directly observable are only
colorless formations of a quark and antiquark (mesons). There is,
of course, no known way to analytically derive confinement from
first principles up to now. Nevertheless some evidence has been
obtained at the level of computer simulations based on lattice
gauge theories \cite{Capstick}.\\
In this context, it may be recalled that many authors have linked
the confinement phenomena to the instability of the perturbative
$QCD$ vacuum \cite{Savvidy,Nielsen}, which indicates the existence
of condensates of gauge fields.
Confinement as a consequence  of the interaction between a
constant chromo-magnetic background and
the axion field has been recently discussed in \cite{pe}.\\
The need for constant condensates of gauge fields appears already
evident in the loop expansion for a $(2+1)$-dimensional gauge
theory \cite{Guend}. This is because in $QCD_3$, being a
superrenormalizable theory, the loop expansion probes deeper and
deepen into the infrared region at every increasing loop order. In
particular, in order to cancel all leading infrared divergences it
becomes necessary to introduce constant gauge field potentials
(which are not pure gauge in the non-Abelian theory) into the
vacuum. An appropriate gaussian averaging  over the field
strengths and directions is required as well (here the gaussian
width in the averaging must be taken to zero to achieve
cancellation of infrared divergences)
\cite{Guend}.\\
On the other hand, in Ref. \cite{Guend-Owen} a qualitative
analysis of the effect of these constant gauge fields was started
and the arguments presented there indicated that the constant
gauge fields induce linear potentials between static charges. In
fact, it was proposed a reformulation of pure $SU(2)$ Yang-Mills
theory in terms of new variables suitable for its low energy
content. In terms of these variables the long distance physics is
described by an effective action which is $U(1)$ gauge theory.
The idea that confinement is basically an \textit{abelian} effect
can be dated back to \cite{luscher}. More recently it has been
related to relativistic membrane dynamics in  \cite{gaba}, and
implemeted through the \textit{abelian projection} method in
\cite{kondo}. Some peculiar quantum aspects of the effective long
range dynamics of QCD, and certain intriguing analogies with the
Schwinger model, has been discussed in  \cite{aaes}.
Clearly, in order to put our qualitative analysis into a firmer
footing, one needs an explicit calculation for the interaction
energy. In this Letter we address this question and show that
$SU(2)$ Yang-Mills theory can be further mapped into the massive
Schwinger model \cite{Schwin, GaeteSch}. Our calculation is based
on the gauge-invariant but path-dependent variables formalism
\cite{Gaete}. According to this formalism, the interaction
potential between two static charges is obtained once a judicious
identification of the physical degrees of freedom is made. It also
offers an alternative technique for determining the static
potential for a gauge theory.
The novel ingredient we introduce in this letter is the
computation of the static potential in the presence of one, or
more, compact spacelike dimension. In this sort of Kaluza-Klein
approach the limit of infinite radius of  compactification can be
performed provided a self-consistency condition between the
ultraviolet cut-off and the amplitude of the constant background
field  is satisfied.

\section{Interaction Energy}

As already stated, our principal purpose is to calculate
explicitly the interaction energy between static pointlike sources
for three-dimensional $SU(2)$ Yang-Mills theory. To this end  we
will calculate the expectation value of the energy operator $ H$
in the physical state $ |\Phi\rangle$, which we will denote by ${
\langle H \rangle_ \Phi}$. However, before going to the derivation
of the interaction energy, we will describe very briefly the new
set of variables which map $SU(2)$ Yang-Mills theory to the
Abelian gauge model. We take as starting point the
three-dimensional space-time Lagrangian:
\begin{equation}
\mathcal{ L} =  - \frac{1}{4}Tr\left( {G_{\mu \nu } G^{\mu \nu } }
\right)= - \frac{1}{4}G_{\mu \nu }^a G^{a\mu \nu }. \label{CFC10}
\end{equation}
Here $ A_\mu  \left( x \right) = A_\mu ^a \left( x \right)T^a $,
where $T^a$ is a hermitian representation of the semi-simple and
compact gauge group; and $ G_{\mu \nu }^a  = \partial _\mu  A_\nu
^a - \partial _\nu  A_\mu ^a  + g\epsilon^{abc} A_\mu ^b A_\nu
^c$, with $\epsilon^{abc}$ the structure constants of the $SU(2)$
gauge group.\\
We should now consider the reformulated version of this theory,
which has been developed and analyzed in \cite{Guend-Owen}, where
one can find the motivations to consider it and details of the
change of variables are given. Applying the change of variables
defined by the fields \cite{Guend-Owen}: $ W_\mu ^ \pm   =
{\raise0.7ex\hbox{$1$} \mathord{\left/ {\vphantom {1 {\sqrt 2
}}}\right.\kern-\nulldelimiterspace} \!\lower0.7ex\hbox{${\sqrt 2
}$}}\left[ {A_\mu ^{\left( 1 \right)}  \pm \left( { - iA_\mu
^{\left( 2 \right)} } \right)} \right]$ and $ A_\mu \equiv
A_{\mu}^{(3)}$, on (\ref{CFC10}), one obtains
\begin{equation}
\mathcal{ L} =  - \frac{1}{4}F_{\mu \nu } F^{\mu \nu }  -
\frac{1}{2}\left| {D_\mu  W_\nu ^ +   - D_\nu  W_\mu ^ -  }
\right|^2  + \frac{g}{2}F_{\mu \nu } W_\rho ^ -  S^{\mu \nu \rho
\sigma } W_\sigma ^ -   - \frac{{g^2 }}{4}W_\rho ^ -  S^{\mu \nu
\rho \sigma } W_\sigma ^ +  W^{ - \alpha } S_{\mu \nu \alpha \beta
} W^{ + \beta } , \label{CFC15}
\end{equation}
where  $F_{\mu \nu }  = \partial _\mu  A_\nu   - \partial _\nu
A_\mu$, $D_\mu   = \partial _\mu   - igA_\mu$ and $S_{\mu \nu \rho
\sigma }  = i\left( {g_{\mu \rho } g_{\nu \sigma }  - g_{\mu
\sigma } g_{\nu \rho } } \right)$. Next, according to Ref.
\cite{Guend-Owen}, we restrict our attention to the ground state
given by:
\begin{equation}
\begin{array}{l}
   \\
                               \left\{ \begin{array}{l}
 A_\mu ^{\left( 1 \right)}  = C\left( constant \right)\delta _\mu^1 \\
 A_2^{\left( 1 \right)}  = A_0^{\left( {1,2} \right)}  = 0 \\
 \end{array} \right\} \Rightarrow \left\langle {W_\mu ^ +  } \right\rangle
= \left\langle {W_\mu ^ -  } \right\rangle  = \frac{C}{{\sqrt 2
}}\delta _{\mu 1}.  \\  \label{CFC20}
 \end{array}
\end{equation}
Hence the Lagrangian (\ref{CFC10}) reduces to:
\begin{equation}
\mathcal{ L} =  - \frac{1}{4}F_{\mu \nu } F^{\mu \nu }  +
\frac{1}{2} m^2 A_\mu A^\mu -A_0 J^0, \label{CFC25}
\end{equation}
where $J^0$ is the external current, and $g^2 C^2\equiv m^2$. In
this Lagrangian we have considered only sources and gauge field
excitations in the third direction of $SU(2)$ only. This appears
to be enough in order to study the basic features of confinement,
as we will show now. Thus the Lagrangian (\ref{CFC10}) becomes a
Maxwell theory supplemented by the Proca term. Notwithstanding, in
order to put our discussion into context it is useful to summarize
the relevant aspects of the canonical quantization of this
reformulated theory from the Hamiltonian point of view.\\
We now proceed to obtain the Hamiltonian. The canonical momenta
read $ \Pi ^{\mu } = - F^{0\mu }$, which results in the usual
primary constraints $ \Pi _0 = 0$, and $ \Pi ^{i}  = F^{i0}$.
Standard techniques for constrained systems then lead to the
following canonical Hamiltonian
\begin{equation}
H_C  = \int {d^2 x} \left\{ { - A_0 \left( {\partial _i \Pi ^i  +
m^2  - J^0 } \right) - \frac{1}{2}\Pi _i \Pi ^i  +
\frac{1}{4}F_{ij} F^{ij}  - m^2 A_i A^i } \right\}. \label{CFC30}
\end{equation}
Requiring the primary constraint to be stationary, we obtain
\begin{equation}
\Gamma  \equiv \partial _i \Pi ^i  + m^2 A^0  - J^0=0.
\label{CFC35}
\end{equation}
It can be easily checked that both constraints are second class.
This result is not surprising, it explicitly reflects the breaking
of the gauge invariance of the theory under consideration.
Consequently, special care has to be exercised since it is the
gauge invariance that generally establish unitarity and
renormalizability in most quantum field theoretical models. To
convert the second class system into first class we will adopt the
procedure described in Refs.\cite{Clovis,Rabin}. In this way the
new system still has the basic features of the original one and
has reobtained the gauge symmetry. As stated in Refs.
\cite{Clovis,Rabin,GaeteGuen}, we enlarge the original phase space
by introducing a canonical pair of fields $\theta$ and $\Pi
_\theta$. It follows, therefore, that a new set of first class
constraints can be defined in this extended space:
\begin{equation}
\Lambda _1  \equiv \Pi _0  + m^2 \theta  = 0, \label{CFC40}
\end{equation}
and
\begin{equation}
\Lambda _2  \equiv \Gamma  + \Pi _\theta   = 0. \label{CFC45}
\end{equation}
It is easy to verify that the new constraints are first class and
in this way restore the gauge symmetry of the theory under
consideration. It is to be observed that the $\theta$ fields only
enlarge the unphysical sector of the total Hilbert space, not
affecting the structure of the physical subspace \cite{Clovis}.
Then, the new effective Lagrangian, after integrating out the
$\theta$ fields, becomes
\begin{equation}
\mathcal{ L}_{eff} = - \frac{1}{4}F_{\mu \nu } \left( {1 +
\frac{{m ^2 }}{\Box}} \right)F^{\mu \nu }- A_0 J^0, \label{CFC50}
\end{equation}
one immediately sees that $SU(2)$ Yang-Mills theory has been
mapped into a $U(1)$ gauge theory.\\
Having established the new effective Lagrangian, we can now
compute the interaction energy. To this end, we first consider the
Hamiltonian framework of this new effective theory. The canonical
momenta read $\Pi ^\mu  =  - \left( {1 +\frac{{m^2 }}{\Box }}
\right)F^{0\mu}$. This yields the usual primary constraint
$\Pi^0=0$, and $\Pi ^i  = - \left( {1 + \frac{{m^2 }}{\Box }}
\right)F^{0i}$. Therefore the canonical Hamiltonian takes the form
\begin{equation}
H_C  = \int {d^2x} \left\{ { - A_0 \left( {\partial _i \Pi ^i  -
J^0 } \right) - \frac{1}{{2}}\Pi _i \left( {1 + \frac{{m ^2 }}{
\Box }} \right)^{ - 1} \Pi ^i + \frac{1}{4}F_{ij} F^{ij}}
\right\}. \label{CFC55}
\end{equation}
Temporal conservation of the primary constraint $\Pi_0$ leads to
the secondary constraint $\Gamma _1 \left( x \right) \equiv
\partial _i \Pi ^i  - J^0  = 0$. It is straightforward to check that
there are no further constraints in the theory. The extended
Hamiltonian that generates translations in time then reads $H =
H_C  + \int d x \left( {c_0 (x)\Pi_0 (x) + c_1 (x)\Gamma _1 (x)}
\right)$, where $c_0(x)$ and $c_1(x)$ are the Lagrange
multipliers. Moreover, it follows from this Hamiltonian that $
\dot{A}_0 \left( x \right) = \left[ {A_0 \left( x \right),H}
\right] = c_0 \left( x \right)$, which is an arbitrary function.
Since $\Pi_0 = 0$, neither $A^0$ nor $\Pi^0$ are of interest in
describing the system and may be discarded from the theory. As a
result, the Hamiltonian becomes
\begin{equation}
H = \int {d^2x} \left\{ { - \frac{1}{{2}}\Pi _i \left( {1 +
\frac{{m ^2 }}{\Box }} \right)^{ - 1} \Pi ^i + \frac{1}{4}F_{ij}
F^{ij} + c^ \prime \left( {\partial _i \Pi ^i  - J^0 } \right)}
\right\}, \label{CFC60}
\end{equation}
where $c^ \prime  \left( x \right) = c_1 \left( x \right) - A_0
\left( x \right)$.

Since there is one first class constraint $\Gamma_1(x)$ (Gauss'
law), we choose one gauge fixing condition that will make the full
set of constraints becomes second class. We choose the gauge
fixing condition to correspond to \cite{Gaete,GaeteGuen}
\begin{equation}
\Gamma _2 \left( x \right) \equiv \int\limits_{C_{\xi x} } {dz^\nu
} A_\nu \left( z \right) \equiv \int\limits_0^1 {d\lambda x^i }
A_i \left( {\lambda x} \right) = 0, \label{CFC65}
\end{equation}
where  $\lambda$  $\left( {0 \le \lambda  \le 1} \right)$ is the
parameter describing the spacelike straight path between the
reference points $ \xi ^k $ and $ x^k $ , on a fixed time slice.
For simplicity we have assumed the reference point $\xi^k=0$. The
choice (\ref{CFC65}) leads to the Poincar\'{e} gauge. With this we
obtain the only nontrivial Dirac bracket
\begin{equation}
\left\{ {A_i \left( x \right),\Pi ^{j} \left( y \right)} \right\}^
*   = \delta _i^j \delta ^{(2)} \left( {x - y}
\right) - \frac{\partial }{{\partial x^i }}\int\limits_0^1
{d\lambda }x^j \delta ^{(2)} \left( {\lambda x - y} \right) .
\label{CFC70}
\end{equation}

We pass now to the calculation of the interaction energy, where a
fermion is localized at the origin ${\bf 0}$ and an antifermion at
$ {\bf y}$. As already mentioned, we will calculate the
expectation value of the energy operator $ H$ in the physical
state $ |\Phi\rangle$, which we will denote by ${ \langle H
\rangle_ \Phi}$ . From our above discussion, we see that ${
\langle H \rangle_ \Phi}$ reads
\begin{equation}
\left\langle H \right\rangle _\Phi   = \left\langle \Phi
\right|\int {d^2 x} \left( { - \frac{1}{2}\Pi _i  \left( {1 +
\frac{{m ^2 }}{\Box }} \right)^{ - 1} \Pi ^{i}  +
\frac{1}{4}F_{ij} F^{ij} } \right)\left| \Phi  \right\rangle.
\label{CFC75}
\end{equation}
Since the fermions are taken to be infinitely massive (static),
this can be further simplified as
\begin{equation}
\left\langle H \right\rangle _\Phi   = \left\langle \Phi
\right|\int {d^2 x} \left( { - \frac{1}{2}\Pi _i \left( {1 -
\frac{{m ^2 }}{\nabla ^2}} \right)^{ - 1}\Pi ^{i} } \right)\left|
\Phi \right\rangle . \label{CFC80}
\end{equation}
Let us also mention here that, as was first established by Dirac
\cite{Dirac}, the physical states $|\Phi\rangle$ correspond to the
gauge invariant ones. It is helpful to recall at this stage that
in the Abelian case $|\Phi\rangle$ may be written as \cite{GaeteB}
\begin{equation}
\left| \Phi  \right\rangle  \equiv \left|\overline \Psi \left(
{\bf y} \right)\Psi \left( {\bf 0} \right)   \right\rangle =
\overline \psi \left( {\bf y} \right)\exp \left(
{ig\int\limits_{\bf 0}^{\bf y} {dz^i A_i \left( z \right)} }
\right)\psi \left( {\bf 0} \right)\left| 0 \right\rangle ,
\label{CFC85}
\end{equation}
where $|0\rangle$ is the physical vacuum state and the line
integral appearing in the above expression is along a spacelike
path starting at $\bf 0$ and ending at $\bf y$, on a fixed time
slice. It is to be observed that the strings between fermions have
been introduced in order to have a gauge-invariant function $
\left| \Phi \right\rangle $. Another way of saying the same thing
is that the fermion fields are now dressed by a cloud of gauge
fields.\\
Returning now to our problem on hand, we compute the expectation
value of $H$ (given by the expression (\ref{CFC80})) in the
physical state $|\Phi\rangle$. Taking into account the above
Hamiltonian structure, we observe that
\begin{equation}
\Pi _i \left( x \right)\left| {\overline \Psi  \left( y
\right)\Psi \left( 0 \right)} \right\rangle  = \overline \Psi
\left( y \right)\Psi \left( 0 \right)\Pi _i \left( x \right)\left|
0 \right\rangle  - e\int_{\bf 0}^{\bf y} {dz_i } \delta ^{\left( 2
\right)} \left( {z  - x} \right)\left| \Phi  \right\rangle.
\label{CFC90}
\end{equation}
Inserting this back into (\ref{CFC80}), we get
\begin{equation}
\left\langle H \right\rangle _\Phi   = \left\langle H
\right\rangle _0  - \frac{e^2 }{2\pi}K_0(mL), \label{CFC95}
\end{equation}
where $\left\langle H \right\rangle _0  = \left\langle 0
\right|H\left| 0 \right\rangle$ and with $|{\bf y}|\equiv L$.
Since the potential is given by the term of the energy which
depends on the separation of the two fermions, from the expression
(\ref{CFC95}) we obtain
\begin{equation}
V= - \frac{e^2 }{2\pi}K_0(mL). \label{CFC100}
\end{equation}
In this way the static interaction between fermions arises only
because of the requirement that the $\left| {\overline \Psi \Psi }
\right\rangle$ states be gauge invariant. It is interesting to
note that this is exactly the result obtained for the
Maxwell-Chern-Simons theory \cite{Abdalla, GaeteD}.\\
We are now in position to examine the mapping of the theory
(\ref{CFC50}) into the massive Schwinger model. Let us illustrate
this by making a dimensional compactification (\`a la
Kaluza-Klein) on Eq.(\ref{CFC50}). The compactification is needed
also in order to control the infrared behavior of the theory. In
such a case, the new theory takes the form:
\begin{equation}
\mathcal{ L}^{(1 + 1)}  =  - \frac{1}{4}F_{\mu \nu } \sum\limits_n
{\left( {1 + \frac{{g^2 C^2 }}{{\Box _{(1 + 1)}  + a^2 }}}
\right)F^{\mu \nu }  - A_0 J^0 }, \label{CFC110}
\end{equation}
where $a^2  \equiv {\raise0.7ex\hbox{${n^2 }$} \!\mathord{\left/
 {\vphantom {{n^2 } {R^2 }}}\right.\kern-\nulldelimiterspace}
\!\lower0.7ex\hbox{${R^2 }$}}$, and $R$ is the compactification
radius. We immediately recognize the above to be the massive
Schwinger model with mass $m^{2} \equiv a^{2}$. We observe, as an
appealing feature of this expression, that for the zero mode
($a=0$) case it gives the massless Schwinger model. It is now once
again straightforward to compute the interaction energy. Taking a
contribution of a single mode in Eq.(\ref{CFC110}), we obtain
\cite{GaeteSch}:
\begin{equation}
V = \frac{{e^2 }}{{2\lambda }}\left( {1 + \frac{{a^2 }}{{\lambda
^2 }}} \right)\left( {1 - e^{ - \lambda L} } \right) + \frac{{e^2
}}{2}\left( {1 - \frac{{g^2 C^2 }}{{\lambda ^2 }}} \right)L,
\label{CFC115}
\end{equation}
where $\lambda ^2  \equiv g^2 C^2  + a^2$ and $|{y}|\equiv L$.
Effectively, therefore, our initial theory (\ref{CFC10}) is mapped
into the massive Schwinger model, which displays both the
screening and the confining part of this interaction. Of course,
if we consider the zero mode case, i. e. , $a=0$, the static
potential above shows that confinement disappears. When studying
this expression, we will concentrate on the second term of Eq.
(\ref{CFC115}), which represents confinement. We start considering
the subtle points related to the calculation of this confinement
term. Although, as we see from Eq. (\ref{CFC115}), every single
mode contribution shows confinement, one can ask the question of
how to handle the sum over all modes in (\ref{CFC115}). If we are
interested in the interaction potential between two point-like
sources in the $(2+1)$-dimensional case (and later we will
consider also the $(3+1)$-dimensional case) we must sum over all
the modes with equal weight, according to the representation of
the delta function source.\\
Basically, the expression for the coefficient of the linear
potential between two static point sources is:
\begin{equation}
T = \frac{{e^2 }}{2}\sum\limits_n {\frac{{{\raise0.7ex\hbox{${n^2
}$} \!\mathord{\left/
 {\vphantom {{n^2 } {R^2 }}}\right.\kern-\nulldelimiterspace}
\!\lower0.7ex\hbox{${R^2 }$}}}}{{g^2 C^2  +
{\raise0.7ex\hbox{${n^2 }$} \!\mathord{\left/
 {\vphantom {{n^2 } {R^2 }}}\right.\kern-\nulldelimiterspace}
\!\lower0.7ex\hbox{${R^2 }$}}}}}. \label{115A}
\end{equation}
In the limit $R \to \infty$, the sum can be approximated by an
integral. Defining the continuous variable $x = \frac{n}{R}$, Eq.
(\ref{115A}) can be rewritten as
\begin{equation}
T = \frac{{e^2 }}{2}R\int_0^\Lambda  {dx\frac{{x^2 }}{{g^2 C^2  +
x^2 }}}, \label{115B}
\end{equation}
where $\Lambda$ is an ultraviolet cutoff. After some
manipulations, we get that $T$ is given by
\begin{equation}
T = \frac{{e^2 R}}{2}\left( {\Lambda  - \frac{\pi }{4}gC} \right)
, \label{115C}
\end{equation}
From expression (\ref{115C}), in the limit $R \to \infty$, the
only way to obtain a finite value for $T$ is to let $C \to \infty$
according to
\begin{equation}
C \to \frac{{4\Lambda }}{{\pi g}} + \frac{\kappa }{R},
\label{115D}
\end{equation}
where $\kappa$ is a constant which determines the value of the
constant $T$. In other words, the external background field
(\ref{CFC20}) has to be selected of the order of the ultraviolet
cutoff. This is in complete accordance with the result of
\cite{Guend}, where when considering the cancellation of infrared
divergences one has to consider constant external gauge fields of
infinite value as the cutoff is removed.\\
At this stage, it is interesting to recall the results reported in
Refs. \cite{Schro, Lavelle} for a $(2+1)$-dimensional $SU(N)$
Yang-Mills theory
\begin{equation}
V(L) = \frac{{g^2 C_F }}{{2\pi }}\log (g^2 L) + \frac{7}{{64\pi }}
g^4 C_F C_A L, \label{CFC120}
\end {equation}
where $C_{F}$ and $C_{A}$ are the Casimir group factor. Hence we
see that our phenomenological result (\ref{CFC115}) agrees
qualitatively with (\ref{CFC120}) in the limit of large $L$.
Corroborating that the reformulation of $SU(2)$ Yang-Mills theory
in terms of the new set of variables is suitable for its low
energy content. In this way the massive Schwinger model simulates
the features of the $SU(2)$ Yang-Mills theory. Notice that in our
teatment, as $C \to \infty$ the first term in (\ref{CFC115}) (even
after summation) does not lead to a non trivial contribution.

\section {Final remarks}

In summary, we have considered the confinement versus screening
issue for a reformulated three-dimensional Yang-Mills theory. This
reformulation is achieved due to the presence of condensates of
gauge fields, which are implemented by a new set of variables. In
terms of these variables the long distance physics is described by
an effective action which is a $U(1)$ gauge theory leading to a
linear potential between static charges.\\
In the $(3+1)$-dimensional case, we can perform the
compactification of two spatial dimensions, assigning to them
radii $R_1$ and $R_2$, we obtain for the coefficient of the linear
potential the expression
\begin{equation}
T = \frac{{e^2 }}{2}\sum\limits_{n_1 ,n_2 } {\frac{{\left(
{\frac{{n_1^2 }}{{R_1^2 }} + \frac{{n_2^2 }}{{R_2^2 }}}
\right)}}{{{ g^2 C^2 } + \frac{{n_1^2 }}{{R_1^2 }} + \frac{{n_2^2
}}{{R_2^2 }}}}}. \label{FR10}
\end{equation}
Following the same steps of the $(2+1)$-dimensional case, in the
limit $R_1,R_2 \to \infty$ we obtain
\begin{equation}
T = \pi e^2 R_1 R_2 \int_0^\Lambda  {d\rho \frac{{\rho ^3 }}{{{C^2
} + \rho ^2 }}}, \label{FR20}
\end{equation}
that is,
\begin{equation}
T = \frac{{\pi e^2 }}{2}R_1 R_2 \left[ {\Lambda ^2  - g^2 C^2 \ln
\left( {\frac{{g^2 C^2  + \Lambda ^2 }}{{g^2 C^2 }}} \right)}
\right], \label{FR30}
\end{equation}
again if $R_1,R_2  \to \infty$, we obtain the transcendental
equation for ${\raise0.7ex\hbox{$\Lambda $} \!\mathord{\left/
 {\vphantom {\Lambda  {C^2 }}}\right.\kern-\nulldelimiterspace}
\!\lower0.7ex\hbox{${g^2 C^2 }$}}$:
\begin{equation}
\frac{{\Lambda ^2 }}{{g^2 C^2 }} - \ln \left( {1 + \frac{{\Lambda
^2 }}{{g^2 C^2 }}} \right) = 0. \label{FR40}
\end{equation}
From (\ref{FR30})  here we can deduce that as $\frac{{\Lambda ^2
}}{{g^2 C^2 }} \sim \sqrt {\frac{{2\pi T}}{{e^2 R_1 R_2 }}} \to
0$, which means that the external field $C$ has to grow stronger
than $g\Lambda$ when $\Lambda  \to \infty$, in order to obtain a
finite coefficient of the linear potential.\\
Although this method appears simple and intuitive, it displays the
basic features of how a confining potential can appear even in the
$(3+1)$-dimensional gauge theory. Finally, notice that once the
space has undergone a compactification, the background field
(\ref{CFC20}) may not be transformed away by a gauge
transformation, since such a gauge transformation in general would
violate the periodicity conditions. Another way of understanding
the physical reality of the background (\ref{CFC20}) is by
calculating the gauge invariant Wilson loops of (\ref{CFC20})
along the compactified dimensions, which are non vanishing.

\section{ACKNOWLEDGMENTS}

One of us (E. G.) wants to thank the Physics Department of the
Universidad T\'{e}cnica F. Santa Mar\'{\i}a for hospitality. P. G.
was partially supported by FONDECYT (Chile) grant 1050546.

\end{document}